\documentclass[preprint,12pt]{elsarticle}

\usepackage{amssymb}
\usepackage{amsmath}
\usepackage{booktabs}    
\usepackage{adjustbox}   
\usepackage{ragged2e}

\begin{document}

\begin{frontmatter}



\title{New level density parameter beyond Egidy-Bucurescu's systematics}


\author{Junzhe Zhang, Yanan Zheng, Caixin Yuan, Yangyang Shen} 
\author[a]{Yingchen Mao} 
\ead{myc@lnnu.edu.cn}

\address{Department of Physics, Liaoning Normal University, Dalian, 116029, PR China}

\begin{abstract}

Extending beyond the Egidy-Bucurescu systematics, the nuclear level density parameters (LDPs) for the back-shifted Fermi gas model were compiled. Three forms of LDPs were fitted: the liquid-drop model (LDM), the droplet model (DM), and the power-law dependence on mass number A. Additionally, the root-mean-square deviations (RMSDs) of the new LDPs and existing literature values were calculated. The newly fitted global LDM-type parameters outperform the commonly used Töke-Swiatecki parameters in various statistical model calculations. In contrast, neither the global nor the combined DM-type parameters yielded satisfactory results. Among the tested parameter sets, the widely adopted Reisdorf parameters exhibited the best overall performance, as evidenced by the larger number of experimental data points falling within their narrower RMSD confidence intervals. For the power-law A-dependence, the new global parameters performed better than the existing homogeneous ones. The ground-state deformation and isospin correction factors had minimal overall impact on the LDP fits. However, the current results suggest that theoretical calculations for transitional nuclei should account for ground-state deformation effects.
\end{abstract}

\end{frontmatter}



\section{Introduction}
\label{sec1}

Nuclear level density (NLD), a fundamental quantity in nuclear reaction models, characterizes the number of single-particle levels in a given energy bin, usually 1 MeV. It plays a critical role in modeling heavy element nucleosynthesis ~\cite{Burbidge1957_RMP, Kajino2019_PPNP, Cowan2021_RMP, Larsen2019_PPNP}, fission dynamics ~\cite{Schunck2022_PPNP, Frobrich1998_PR, Schmidt2018_RPP}, and reaction cross section calculations ~\cite{Back2014_RMP, Hauser1952_PR}. At low energies, the NLDs are discrete and sparse. With increasing excitation energy, the single-particle levels gradually merge into quasi-continuous states; on the other hand, the spacings of levels will greatly reduce. In the simplest form, the NLD can be evaluated by treating the excited nucleus as non-interacting fermions confined in the finite nuclear volume and with equally spaced energy. Based on the Fermi gas model, Bethe conducted the first theoretical model to calculate the NLD ~\cite{Bethe1936_PR}, but the model yields unreliable predictions. To enhance the reliability of model predictions, various theoretical models have been progressively developed to effectively characterize NLD, such as the Gilbert-Cameron model ~\cite{Gilbert1965_CJP}, the generalized superfluid model ~\cite{Ignatyuk1979_SJNP, Ignatyuk1993_PRC}, the statistical partition function ~\cite{Sepiani2023_JPG, Williams1969_NPA, Mainegra2003_CPC}, the shell model Monte Carlo method ~\cite{Nakada1997_PRL, Alhassid1999_PRL, Alhassid2007_PRL, Koonin1997_PR}, the projected shell model ~\cite{Wang2023_PRC}, the relativistic density functional theory ~\cite{Jiang2024_PLB}, the stochastic estimation method ~\cite{Shimizu2016_PLB, Chen2023_PRC} and the extrapolated Lanczos matrix approach ~\cite{Ormand2020_PRC}, etc. 

Although significant progress has been made in theoretical descriptions of NLD, the most widely used model in statistical calculations is the back-shifted Fermi gas (BSFG) model ~\cite{Gilbert1965_CJP, Brancazio1969_CJP} rooted in Bethe's foundational work ~\cite{Bethe1936_PR}, and energy-dependent BSFG model ~\cite{Ignatyuk1975_SJNP, Iljinov1992_NPA, Rauscher1997_PRC}, which takes the shell correction and damping effect. In the BSFG framework, there is a key parameter serve as dominance role, namely \textit{a}, nuclear level density parameter (LDP). With the LDP \textit{a}, the NLD can be simply approximated as
\begin{equation}
\rho  \propto (2L + 1){e^{2\sqrt {a{E^*}} }}
\end{equation}
where \textit{L} and \textit{E*} are angular momentum and excitation energy of the nucleus, respectively. In early statistical models, such as the ALICE code ~\cite{Blann1984_PRC}, the PACE2 code ~\cite{Gavron1980_PRC} and the GEMINI code ~\cite{Charity1988_NPA}, \textit{a} is treated as \textit{A}/\textit{k}, where \textit{A} is the mass number of the nucleus and \textit{k} is a constant. In this situation, a corresponds as the contribution of the Fermi gas’s volume term, and the relationship between \textit{a} and the single-particle level \textit{g} near Fermi energy $\epsilon_{f}$ is
\begin{equation}
a = \frac{{{\pi ^2}}}{6}g({\varepsilon _f})
\end{equation}
Considering the inhomogeneity and finite boundaries of real nuclear systems, the single-particle level \textit{g} should take into account some corrections, namely, surface corrections, curvature corrections, and others. Ultimately, the LDP $a$ can be expressed as
\begin{equation}
a = {a_{\rm{v}}}A + {a_{\rm{s}}}{A^{{2 \mathord{\left/
 {\vphantom {2 3}} \right.
 \kern-\nulldelimiterspace} 3}}}{B_{\rm{s}}}
 \end{equation}
 or \begin{equation}a = {a_{\rm{v}}}A + {a_{\rm{s}}}{A^{{2 \mathord{\left/
 {\vphantom {2 3}} \right.
 \kern-\nulldelimiterspace} 3}}}{B_{\rm{s}}} + {a_{\rm{k}}}{A^{{1 \mathord{\left/
 {\vphantom {1 3}} \right.
 \kern-\nulldelimiterspace} 3}}}{B_{\rm{k}}}
 \end{equation}
these two forms coincide with the liquid-drop model and the droplet model, ${a_{\rm{v}}}$, ${a_{\rm{s}}}$ and ${a_{\rm{k}}}$ are volume energy, surface energy and curvature energy coefficients, respectively. ${B_{\rm{s}}}$ and ${B_{\rm{k}}}$ are the shape factors of the surface energy and the curvature energy, respectively. 
 
Besides that, a phenomenological expansion of the LDP $a$ in terms of \textit{A} usually can be written as
\begin{equation}
a = {a_1}A + {a_2}{A^2}
\end{equation}
${a_{\rm{1}}}$ and ${a_{\rm{2}}}$ are undetermined coefficients. 

For the BSFG model, with different degrees of correction, there are various different parameters. According to incomplete statistics, there are at least dozens of LDP parameters in the literature, which caused great challenges to the calculation of statistical models. About 20 years ago, von Egidy and Bucurescu provided the LDP corresponding to the BSFG model based on experimental data from 310 nuclide NLDs ~\cite{Egidy2005_PRC, Egidy2009_PRC}, which provided a basis for selecting appropriate model parameters for statistical model calculations of unknown nuclides.

Experimentally, the method of the NLD measurement in the low-energy region is to directly count the low-lying bound levels, while in the high-energy region, it is often achieved by measuring the neutron resonance spacing of stable nuclei or their nearest atomic nuclei near the neutron separation energy from several reactions [(n,$\gamma$), (p, $\gamma$), (p, $p'$), (p, $\alpha$), ($\alpha$, $\gamma$), ($\alpha$, n), (d, p),(${}^3{\rm{He}}$, d),\quad(${}^3{\rm{He}}$, t), (${}^3{\rm{He}}$, ${}^3{\rm{He}}$), (${}^3{\rm{He}}$, $\alpha$)]. Over the past 20 years, the Oslo research team has utilized their own developed methods, namely, the Oslo method ~\cite{Schiller2000_NIMA, Schiller2000_PRC, Olso database2025}, to simultaneously extract the NLDs and $\gamma$-ray strength functions of over 100 nuclides from the primary $\gamma$ spectra close to neutron binding energy ~\cite{Schiller2000_NIMA, Schiller2000_PRC, Olso database2025}, many of which are not included in Refs.~\cite{Egidy2005_PRC, Egidy2009_PRC} Very recently, NLDs have also been obtained using a particle evaporation technique, namely proton evaporation spectra from lithium-induced reactions ($^{6, 7} {\rm{Li}}$, \textit{x}p) ~\cite{Voinov2024_PRC}. Therefore, beyond the works of von Egidy and Bucurescu, we further collect LDP data corresponding to the BSFG model in this paper, and carry out refitting the phenomenological formula parameters of LDP to the new dataset, in order to provide more reliable suggestion and reference for the future statistical model calculations.

\section{Theoretical methods}
\label{sec2}

To perform LDP parameter fitting, we first need to construct an experimental dataset. After a systematic review of the literature, we
have compiled LDP experimental value data for the BSFG model since the work of von Egidy and Bucurescu ~\cite{Egidy2005_PRC}, and also included
LDP experimental data obtained using the Olso method. Finally, we obtained the dataset corresponding to the BSFG models. 

Based on a systematic review of the literature, we also found that the empirical formulas for calculating LDP can generally be divided 
into two categories: one is in the form of droplet 
models (LDM) or small droplet models (DM) expanded 
to a power of ${\textit{A}^{1/3}}$, such as Eqs. (3) and (4), and the other is in the form of droplet models expanded to a power of $\textit{A}$, namely, Eq. (5).

In Eqs. (3) and (4), ${\textit{a}_i} \, (i = {\rm{v\,, s\,, k}})$  corresponds to the volume energy, surface energy, and curvature energy coefficient in LDM (or DM), respectively. The relationship between deformation factor ${\alpha _2}$ and fourth-order deformation parameter ${\beta _2}$ can be expressed as
\begin{eqnarray}
 {\alpha _2} = \sqrt {4/5\pi } {\beta _2}.
\end{eqnarray}
${\textit{B}_s}$ and ${\textit{B}_k}$ are the shape factors of surface energy and curvature energy~\cite{Hasse1988_book, Myers1974_AP}, respectively 

\begin{eqnarray}
 {B_{\rm{s}}} = 1 + \frac{2}{5}\alpha _2^2 - \frac{4}{{105}}\alpha _2^3 - \frac{{66}}{{175}}\alpha _2^4,\\
{B_{\rm{k}}} = 1 + \frac{2}{5}\alpha _2^2 - \frac{{16}}{{105}}\alpha _2^3 - \frac{{82}}{{175}}\alpha _2^4.
\end{eqnarray}

The second type of formula can be expressed as Eq. (5), where $a_1$ and $a_2$ are the coefficients of $\textit{A}$ and ${\textit{A}^{2}}$, respectively.

\section{New parameter fitting based on the backward shifted Fermi gas (BSFG) model}

\subsection{${\textit{A}^{1/3}}$ power related parameter fitting}

In the process of literature research, we have compiled the coefficients of the existing LDM-type LDP and the DM-type LDP in Table 1.

\begin{table}
\centering
\caption{LDM and DM empirical formula coefficient values corresponding to different literature}
\vspace{10pt}
\label{tab:coefficients1}
\begin{adjustbox}{max width=\textwidth} 
\begin{tabular}{l r r r r l c l r r r r l} 
\toprule
\multicolumn{6}{c}{Coefficients} &  & \multicolumn{6}{c}{Coefficients} \\ 
\cmidrule(lr){2-5} \cmidrule(lr){9-12} %
 $N^{\text{a)}}$ & $a_v$ & $a_s$ & $a_k$ & $r_0$ & Refs. & & $N^{\text{a)}}$ & $a_v$ & $a_s$ & $a_k$ & $r_0$ & Refs. \\
 \midrule
1  & 0.073  & 0.095    &       & 1.153  & \cite{Ignatyuk1975_YZ}  & & 15 & 0.054    & 0.233 & 0.103    & 1.153  & \cite{Barranco1981_NPA} \\
2  & 0.055 & 0.150    &       & 1.153  & \cite{Sauer1976_NPA}  & & 16 & 0.053    & 0.228 & 0.175    & 1.153  &      \\
3  & 0.068 & 0.213 &       & 1.16   & \cite{Toke1981_NPA} & & 17 & 0.053    & 0.139 & 0.267    & 1.153  &      \\
4  & 0.096 & 0.147    &       &        & \cite{Manailescu2014_arXiv}  & & 18 & 0.026    & 0.224 & 0.039    & 1.153  &      \\
5  & 0.114  & 0.098 &       &        & \cite{Iljinov1992_NPA}  & & 19 & 0.068    & 0.218 & 0.166    & 1.153  &      \\
6  & 0.074 & 0.070 &       &        & \cite{Kataria1978_PRC}  & & 20 & 0.054    & 0.287 & 0.009    & 1.153  &      \\
7  & 0.078 & 0.115 &       &        & \cite{Schmidt2016_NDS}  & & 21 & 0.076 & 0.180 & 0.157    & 1.2039 & \cite{Mughabghab1998_PRC} \\
8  & 0.117  & 0.077    &       &        & \cite{Iljinov1992_NPA539}  & & 22 & 0.073 & 0.138 & 0.157 & 1.2039 &      \\
9  & 0.114 & 0.162 &       &        & \cite{Ignatyuk1985_IAEA report}  & & 23 & 0.064 & 0.234 & 0.157    & 1.2039   &      \\
10 & 0.067  & 0.259 &       &        & \cite{Koning2012_NDS}  & & 24 & 0.064 & 0.243 & 0.702 & 1.12     & \cite{Prakash1983_PLB} \\
11 & 0.068    & 0.213  & 0.385 & 1.16   & \cite{Toke1981_NPA}  & & 25 & 0.045 & 0.256 & 0.707 & 1.12     &      \\
12 & 0.069 & 0.179 & 0.164 & 1.15   & \cite{Reisdorf1981_ZPA}  & & 26 & 0.045    & 0.453 & 0.651    & 1.12     &      \\
13 & 0.070   & 0.180 & 0.164 & 1.153  &       & & 27 & 0.092    & 0.036 & 0.275    & 1.153     & \cite{Nerlo-Pomorska2006_PRC} \\
14 & 0.064 & 0.174 & 0.148 & 1.153  & \cite{Shlomo1992_NPA}  & & 28 & 0.098 & 0.253 & 0.253    & 1.153     & \cite{Ignatyuk2018_PRC} \\
\bottomrule
\end{tabular}
\end{adjustbox}
\end{table}

After fitting the dataset of LDP in the appendix, which includes 310 nuclei listed in Ref.~\cite{Egidy2005_PRC} and the new experimental data using the least squares method, the global parameters of LDM-type LDP are given, namely:
\begin{eqnarray}
 a = 0.045A + 0.322{A^{2/3}}{B_{\rm{s}}} .
\end{eqnarray}

Considering the relationship between LDP and mass number mentioned earlier, that is, the law of $a \propto A$, we further divided the mass number \textit{A} into three intervals for parameter fitting, and obtained the combined parameters of LDP, namely:
\begin{eqnarray}
 a = \left\{ \begin{array}{l}
{\rm{  }}0.102A + 0.073{A^{2/3}}{B_{\rm{s}}},   A \in [18,100]\\
 - 0.028A + 0.711{A^{2/3}}{B_{\rm{s}}},   A \in (100,180]\\
{\rm{  }}0.102A - 0.019{A^{2/3}}{B_{\rm{s}}},  A \in (180,252] .
\end{array} \right.
\end{eqnarray}

Using the different parameters of the LDP listed in Table 1, as well as the newly fitted global and combined parameters, we calculated the corresponding LDP values for LDM classes. In order to obtain the global predictive ability of these parameters, we introduced the standard deviation (RMSD) $\sigma$ of LDP to measure the quality of the fitting results, which is defined as:
\begin{eqnarray}
 \sigma  = \sqrt {\frac{1}{{N - t}}\sum {{{({a_{\exp }} - {a_{{\rm{th}}}})}^2}} },
\end{eqnarray}
where, ${a_{\exp }}$ and ${a_{{\rm{th}}}}$ are the experimental values and empirical formula calculated values of LDP, respectively. \textit{N} and \textit{t} are the experimental data and the number of fitting parameters required, respectively. The standard deviation $\sigma$ of two new parameters and a total of 12 parameters listed in Table 1 are shown in Fig. 1.

\begin{figure}[t]
\centering
\includegraphics[width=0.9\textwidth]{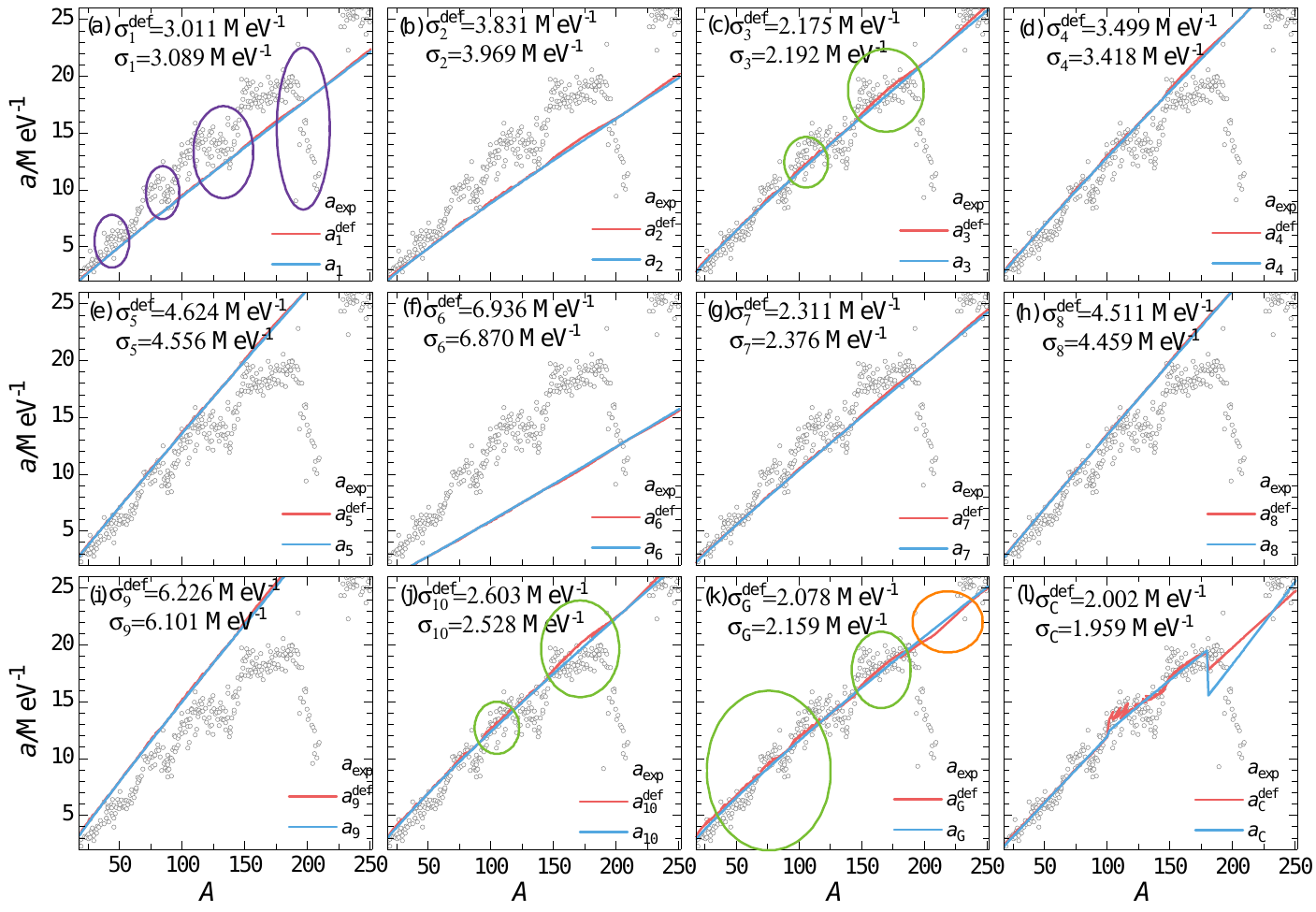} 
\caption{The root-mean standard deviation (RMSD) of empirical formula for the LDM-type LDP}\label{fig1}
\end{figure}

To further investigate the influence of ground-state deformation (represented by shape factors) in empirical formulas such as LDM on the fitting of LDP parameters. We also performed calculations and parameter fitting without considering the ground state deformation [($i.e.~  {B_{\rm{s}}} = {B_{\rm{k}}} = 1$)], and extracted the influence of ground state deformation on parameter fitting by comparing the calculated values of the two cases.

Firstly, we calculated 10 sets of parameters for the LDM class, and then fitted the dataset of LDP in the appendix to obtain new global and combined parameters, which are
\begin{eqnarray}
 a = 0.057A + 0.265{A^{2/3}},
\end{eqnarray}
\begin{eqnarray}
a = \left\{ \begin{array}{l}
0.110A + 0.038{A^{2/3}},   A \in [18,100]\\
0.042A + 0.366{A^{2/3}},  A \in (100,180]\\
0.245A - 0.884{A^{2/3}},  A \in (180,252] .
\end{array} \right.
\end{eqnarray}

From Fig. 1, we can see that the results are better for ${a_{3}}$ and ${a_{G}}$. Compared with the experimental data of LDP, the four sets of parameters, ${a_{1}}$, ${a_{2}}$, ${a_{6}}$ and ${a_{7}}$ significantly underestimate LDP, while the five sets of parameters, ${a_{4}}$, ${a_{5}}$, ${a_{8}}$,${a_{9}}$ and ${a_{10}}$, significantly overestimate LDP. From Fig. 1 (l), it can also be seen that the curve has obvious twists and turns, which is due to the situation where a certain nuclide corresponds to several sets of experimental values during the construction of the dataset. So, although the newly fitted combination formula has the smallest standard deviation value, it is not suitable in terms of parameter fitting effect. Comparing the RMSD values of each group, for the LDP parameters of LDM class, we found that the newly fitted global parameters were better than the existing coefficient values of the third group, namely the Töke-Swiatecki parameter ~\cite{Toke1981_NPA}, which is commonly used in various statistical model calculations. The RMSD values of the global and Töke-Swiatecki parameters are ${\sigma _{\rm{G}}} = 2.078~ {\rm{Me}}{{\rm{V}}^{ - 1}}$ and ${\sigma _3} = 2.175~ {\rm{Me}}{{\rm{V}}^{ - 1}}$ , respectively.

We can also observe an interesting phenomenon from the experimental values in Fig. 1, that is, although LDP exhibits an overall pattern of $a \propto A$, the LDP near mass \textit{A} of 28, 78, 125, and 208 is significantly lower than other shell structures (as indicated in the region labeled in Fig. 1 (k)). Therefore, when calculating the statistical model of nuclides near mass \textit{A} of 208, we recommend selecting a relatively smaller LDP value, such as ${a_{6}}$ (as shown in Fig. 1 (f)).

We can also see that the influence of ground state deformation on the fitting of parameters other than ${a_{3}}$, ${a_{10}}$ and ${a_{G}}$ is not significant. However, for ${a_{3}}$ and ${a_{G}}$ (Figs. 1 (c) and (j)), within the mass intervals $A \in \left[ {94,~122} \right]$ and $A \in \left[ {148,~194} \right]$, the LDP considering ground state deformation is greater than the result without considering ground state deformation. For global parameters (Figure 1 (k)), the LDP considering ground state deformation is greater than the result without considering ground state deformation in the mass intervals $A \in \left[ {18,~131} \right]$ and $A \in \left[ {151,~187} \right]$, while the opposite is true in the mass intervals $A \in \left[ {188,~252} \right]$. Overall, for nuclides in the transition zone ($A \in \left[ {60,~150} \right]$ and $A \in \left[ {190,~220} \right]$), the influence of ground state deformation on LDP parameter fitting is relatively significant.

Similar to the fitting method for empirical formulas of LDM class, we provide the global parameters and combined parameters of DM class separately, namely:
\begin{eqnarray}
 a =  - 0.140A + 2.087{A^{2/3}}{B_{\rm{s}}} - 4.244{A^{1/3}}{B_{\rm{k}}} .
\end{eqnarray}
\begin{eqnarray}
a = \left\{ \begin{array}{l}
 - 0.145A + 1.982{A^{2/3}}{B_{\rm{s}}} - 3.722{A^{1/3}}{B_{\rm{k}}},     A \in [18,100]\\
 - 0.171A + 1.991{A^{2/3}}{B_{\rm{s}}} - 2.937{A^{1/3}}{B_{\rm{k}}},     A \in (100,180]\\
 - 0.678A + 8.572{A^{2/3}}{B_{\rm{s}}} - 24.510{A^{1/3}}{B_{\rm{k}}},  A \in (180,252] .
\end{array} \right.
\end{eqnarray}

Based on the existing coefficient values and the newly fitted global and combination coefficient values, we also calculated the nuclear energy density parameters and corresponding standard deviation RMSD (Eq. (11)) for the DM class. The specific values are shown in Fig. 2.

Similar to the LDM empirical formula, we provide global and combined parameters for the DM empirical formula that do not consider ground state deformation

\begin{eqnarray}
a = 0.018A + 0.663{A^{2/3}} - 0.975{A^{1/3}},
\end{eqnarray}
\begin{eqnarray}
a = \left\{ \begin{array}{l}
0.120A - 0.038{A^{2/3}} + 0.145{A^{1/3}},            A \in [18,100]\\
0.762A - 7.087{A^{2/3}} + 19.250{A^{1/3}},          A \in (100,180]\\
9.792A - 115.20{A^{2/3}} + 341.69{A^{1/3}},        A \in (180,252] .
\end{array} \right.
\end{eqnarray}

The influence of ground-state deformation on the fitting of DM class parameters is also shown in Fig. 2. As shown in Fig. 2, the five sets of parameters, ${a_{12}}$, ${a_{15}}$, ${a_{16}}$, ${a_{20}}$, and ${a_{22}}$, agree well with the experimental data. The four sets of parameters, ${a_{14}}$, ${a_{17}}$, ${a_{18}}$ and ${a_{27}}$, significantly underestimate the parameter values, while the eight sets of parameters, ${a_{11}}$, ${a_{19}}$, ${a_{21}}$, ${a_{23}}$, ${a_{24}}$, ${a_{25}}$, ${a_{26}}$ and ${a_{28}}$, significantly overestimate the parameter values. In addition, we can also find that for the LDP empirical formula of DM class, the calculation results of the newly fitted global parameters and combination parameters are not ideal, and the fitting curves of the two sets of parameters show obvious "fluctuations". The RMSD value calculated from the global parameters is ${\sigma _G} = 2.31~{\rm{Me}}{{\rm{V}}^{ - 1}}$, while the RMSD value of the combination parameters is as high as ${\sigma _{\rm{C}}} = 4.641~{\rm{Me}}{{\rm{V}}^{ - 1}}$.

From Fig. 2, it can be seen that the RMSD values calculated for the five sets of parameters, ${a_{12}}$, ${a_{15}}$, ${a_{16}}$, ${a_{20}}$ and ${a_{22}}$, are very close. In order to further compare which of the five sets of parameters is more in line with the experimental values, we added a ±5\% confidence interval to the calculated values of ${a_{12}}$, ${a_{15}}$, ${a_{16}}$, ${a_{20}}$ and ${a_{22}}$ (shaded areas in Figs. 2 (b), (e), (f), (j), and (i)), and then counted the number of experimental data in the above confidence intervals one by one. The corresponding values for the five sets of parameters, ${a_{12}}$, ${a_{15}}$, ${a_{16}}$, ${a_{20}}$ and ${a_{22}}$, were 132, 95, 99, 129, and 117, respectively. So although the standard deviation corresponding to ${a_{12}}$ is not the smallest, the overall effect of the calculated parameter values should be relatively the best, which is the Reisdorf parameter commonly used in the synthesis of superheavy nuclides using LDP ~\cite{Reisdorf1981_ZPA}. In addition, ${a_{20}}$ is relatively good, although the RMSD value of ${a_{16}}$ is the smallest, there are relatively few experimental points within the confidence interval of its calculated value. This is also due to the "significant deviation" of LDP in several quality regions (such as \textit{A}=208).

We can also see that whether to consider the ground state deformation has little effect on the parameters ${a_{27}}$ and ${a_{28}}$, but has a significant impact on other parameters as well as newly fitted global and combined parameters. Unlike LDM parameters, in the mass ranges $A \in \left[ {92,~122} \right]$ and $A \in \left[ {146,~196} \right]$, the results considering ground state deformation are slightly larger than those without considering ground state deformation.

For the newly fitted global parameters and combination parameters (Figs. 2 (s) and (t)), the calculation results without considering ground state deformation are better than those considering ground state deformation, corresponding to smaller $\sigma$ values and smaller deviations from experimental parameter values. But what is very interesting is that from the perspective of data fitting, whether to consider ground state deformation in the transition region has a relatively large impact on the parameter fitting results, which also reflects the need to consider the influence of ground state deformation when fitting LDP parameters for nuclides in the transition region.

\begin{figure}[t]
\centering
\includegraphics[width=0.8\textwidth]{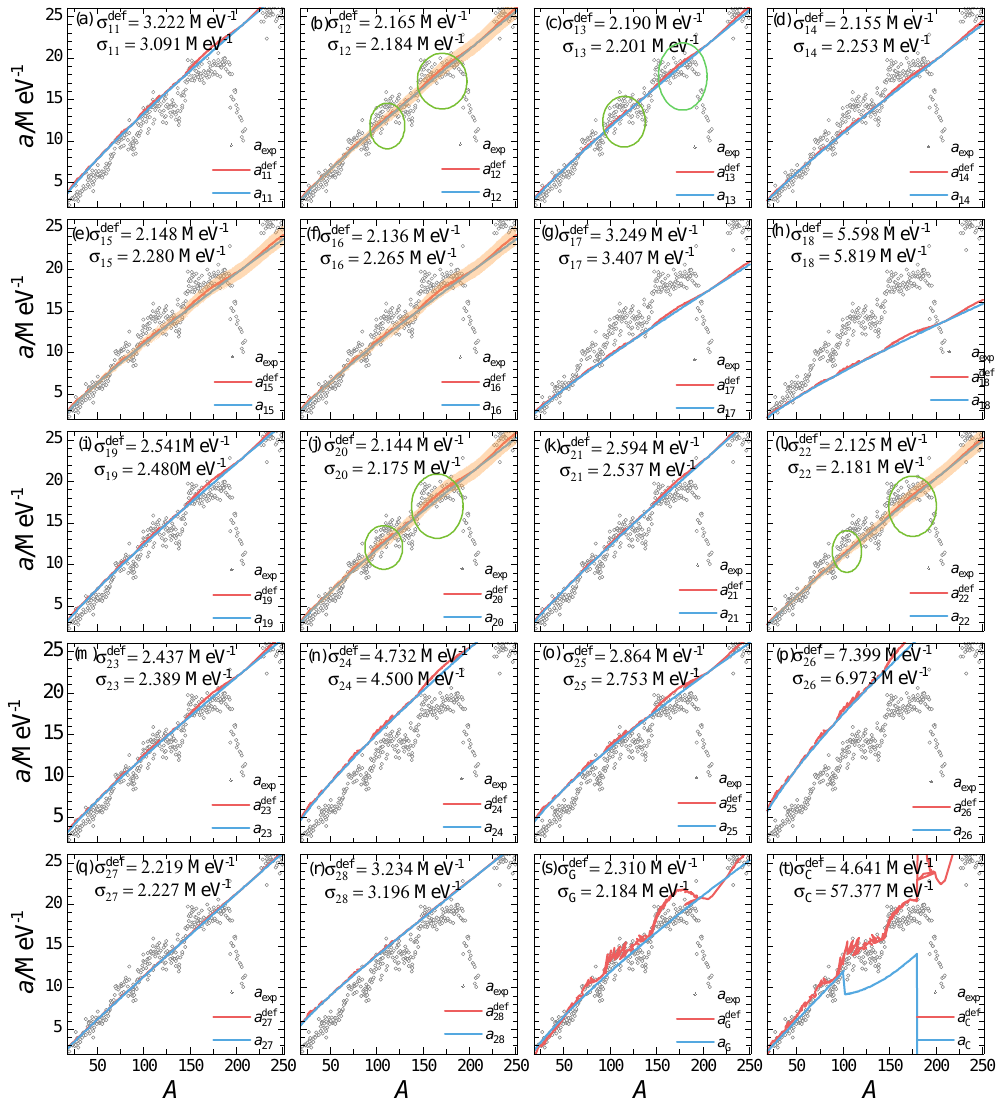} 
\caption{Similar to Fig. 1, the RMSD of the DM-type LDP}\label{fig2}
\end{figure}

Töke and \'{S}wiateck studied the effect of isospin correction factors \[\tilde a = {a_{{\rm{BSFG}}}}(1 - {{{I^2}} \mathord{\left/{\vphantom {{{I^2}} {9)}}} \right.\kern-\nulldelimiterspace} {9)}}\] on LDP parameter~\cite{Toke1981_NPA}, which prompted us to also examine whether isospin correction factors have additional effects on parameter fitting of new datasets. For this purpose, we fitted the LDM class LDP parameters ${a_{3}}$ and DM class LDP parameters ${a_{11}}$ with isospin correction factors included in the corresponding formulas. The new parameters corresponding to ${a_{3}}$ and ${a_{11}}$ are shown in Table 2, and their respective RMSDs are shown in Fig. 3. From Fig. 3, we can see that considering the isospin parameter has almost no effect on the coefficient fitting results.

\begin{table}
\centering
\caption{contains the new parameters of  and  with isospin correction factors}
\vspace{10pt}
\label{tab:coefficients2}
\begin{adjustbox}{max width=\textwidth} 
\begin{tabular}{l r l c l r r l} 
\toprule
\multicolumn{3}{c}{Coefficients} &  & \multicolumn{3}{c}{Coefficients} \\ 
\cmidrule(lr){2-3} \cmidrule(lr){6-8} 
 $N$ & $a_v$ & $a_s$ & & $N$ & $a_v$ & $a_s$ & $a_k$  \\
 \midrule
$a_G^3$  & 0.0702  & 0.210  & & $a_G^{11}$ & 0.0701    & 0.210 & 0.380 \\
$a_c^{\text{a)}}$  & 0.0696 & 0.212  & & $a_c^{\text{a)}}$ & 0.0680   & 0.224 & 0.357 \\
$a_c^{\text{b)}}$ & 0.0707 & 0.207  & & $a_c^{\text{b)}}$ & 0.0694   & 0.219 & 0.351 \\
$a_c^{\text{c)}}$ & 0.0709 & 0.205  & & $ a_c^{\text{c)}}$ & 0.0689    & 0.228 & 0.315 \\
\bottomrule
\end{tabular}
\end{adjustbox}
\end{table} 

\begin{figure}[t]
\centering
\includegraphics[width=0.6\textwidth]{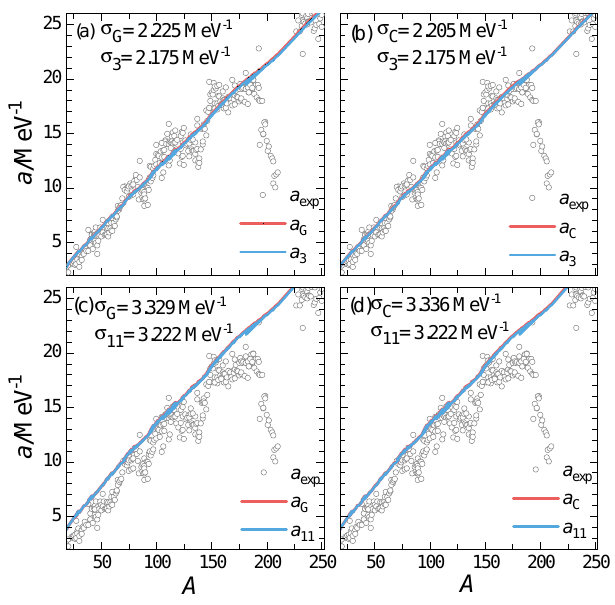} 
\caption{The RMSD of the LDPs with and without considering isospin parameters}\label{fig3}
\end{figure}

\subsection{\textit{A}'s power class parameter fitting} 

Similar to the fitting of LDP parameters for LDM class (or DM class), we have also compiled the parameters for LDP of integer power type \textit{A} and listed them in Table 3. Then, the least squares method was used to fit the parameters of the two datasets, and the corresponding global parameters and combination parameters were represented as

\begin{eqnarray}
 a = 0.125A + 1.10 \times {10^{ - 4}}{A^2},
\end{eqnarray}
\begin{eqnarray}
 a = \left\{ \begin{array}{l}
0.124A - 5.11 \times {10^{ - 5}}{A^2},   A \in [18,100]\\
0.136A - 1.60 \times {10^{ - 4}}{A^2},   A \in (100,180]\\
0.045A + 2.43 \times {10^{ - 4}}{A^2},   A \in (180,252].
\end{array} \right.
\end{eqnarray}

\begin{table}
\centering
\caption{Coefficient values of power-law empirical formulas for A corresponding to different literature}
\vspace{10pt}
\label{tab:coefficients3}
\begin{adjustbox}{max width=\textwidth} 
\begin{tabular}{l r r l c l r r l} 
\toprule
\multicolumn{4}{c}{Coefficients} &  & \multicolumn{4}{c}{Coefficients} \\ 
\cmidrule(lr){2-3} \cmidrule(lr){7-8} 
$N^{\text{a)}}$ & $a_v$ & $a_s$ & Refs. & & $N^{\text{a)}}$ & $a_v$ & $a_s$ & Refs.  \\
\midrule
29 & 0.127  & $9.05 \cdot 10^{-5}$ & \cite{Egidy2005_PRC} & & 32 & 0.126 & $ 7.52 \cdot 10^{-5}$ & \cite{Talou2021_CPC} \\
30 & 0.154 & $6.3 \cdot 10^{-5}$ & \cite{Iljinov1992_NPA} & & 33 & 0.134  & $ -1.21 \cdot 10^{-4}$ & \cite{Iljinov1980_SJNP} \\
31 & 0.138 & $ -8.36 \cdot 10^{-5}$ & \cite{Young1989_LANL report} & & 34 & 0.116  & $8.7 \cdot 10^{-5}$ & \cite{Kawano2006_JNST} \\
\bottomrule
\end{tabular}
\end{adjustbox}
\end{table}

Based on the different parameters given in Table 3 and the new parameters fitted by us, we calculated the integer power class of \textit{A} and its RMSD, and presented them in Fig. 4.

\begin{figure}[t]
\centering
\includegraphics[width=0.8\textwidth]{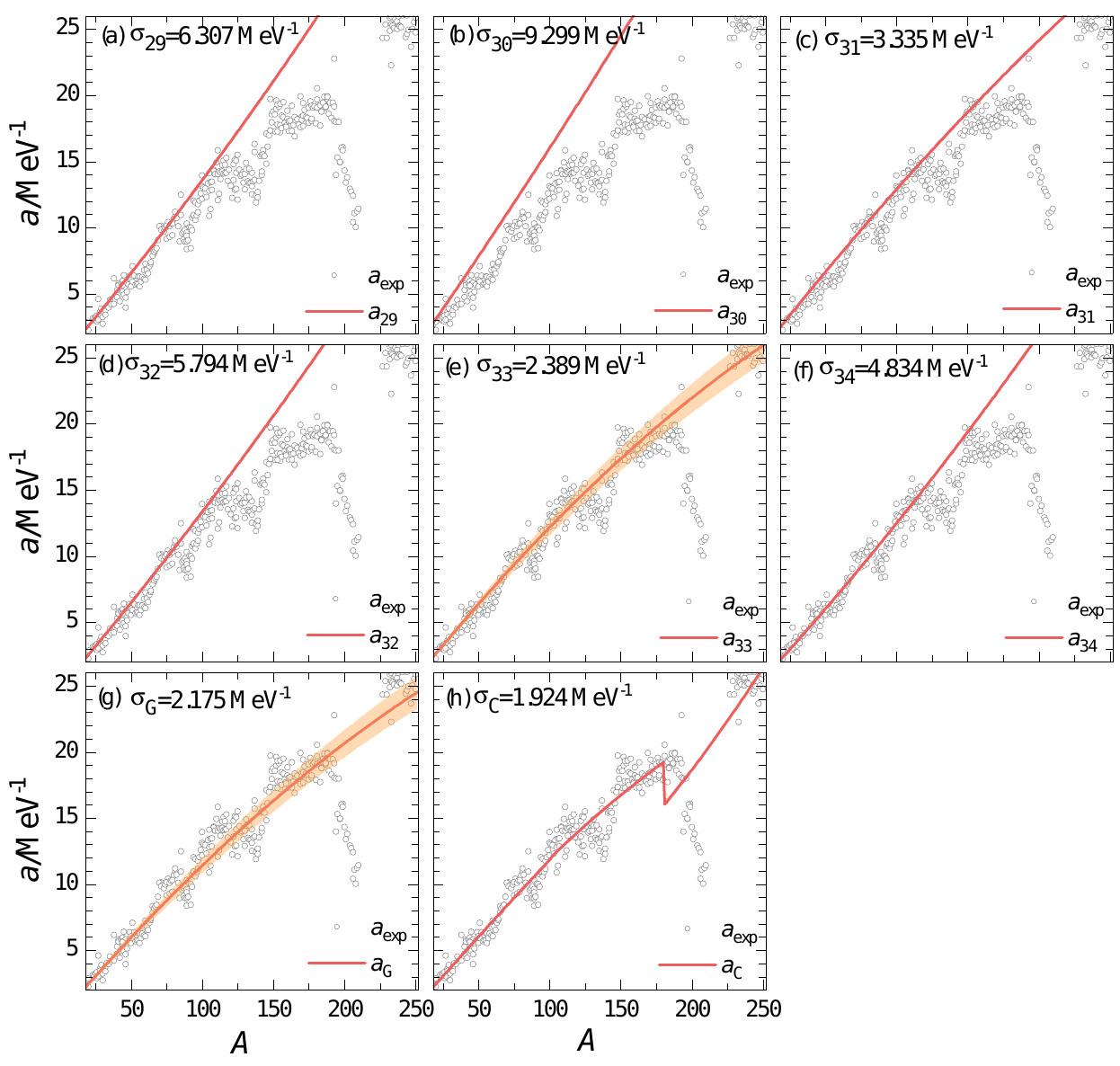} 
\caption{The RMSD of the empirical formula for the $A$-power LDP}\label{fig4}
\end{figure}

From Fig. 4, we can see that the results are better for parameter ${a_{33}}$ and global parameter ${a_{G}}$, while other parameter values significantly overestimate the experimental values. Similarly, we added a ±5\% confidence intervals to the calculated values of ${a_{33}}$ and ${a_{G}}$ (shaded areas in Figs. 4 (e) and (g)), and counted the number of experimental data within the confidence intervals, which were 126 and 121, respectively. The conclusion of adding confidence intervals is the same as that of DM class, that is, although the standard deviation corresponding to ${a_{33}}$ is not the smallest, the overall effect of the calculated parameter values should be relatively the best. Therefore, comparing the RMSD values of each group, we found that ${a_{33}}$ is better than the newly fitted global parameters, with a standard deviation of ${\sigma _{\rm{G}}} = 2.175  {\rm{Me}}{{\rm{V}}^{ - 1}}$ and $a \propto A$ for the global parameters.

In Fig. 4 (h), it can be seen that although the RMSD value of the combined parameters is the smallest, due to the occurrence of several a values corresponding to a certain nuclide during the dataset construction process, the fitting curve shows a significant "bend", which is not suitable from the perspective of parameter fitting effect.

\section{Summary}
\label{sec4}

The LDP values of ${\textit{A}^{1/3}}$ power related droplet models (LDM), small droplet models (DM), and ${\textit{A}^{x}}$ (generally, the highest power of \textit{x} is 2) were fitted using the least squares method for the complete set of mass numbers \textit{A} and three intervals of mass numbers $A \in \left[ {18,100} \right]$,$A \in \left( {100,180} \right]$ and $A \in \left( {180,252} \right]$, respectively. The global and combined parameters were obtained. By comparing the standard deviation (RMSD) of the newly fitted LDP parameters with the RMSD of LDP in the literature, it was found that for LDM type parameter fitting, the global parameters were better than the Töke-Swiatecki commonly used in various statistical model calculations. Parameters. When calculating the statistical model of nuclides with a mass number around 208, we recommend selecting LDP values that are relatively smaller. For DM parameter fitting, the calculation results of the newly fitted global and combined parameters are not ideal. After adding a ±5\% confidence interval to several sets of parameters that match the experimental values well, it was found that the overall effect of using LDP's Reisdorf parameter values frequently should be relatively the best. For parameter fitting of \textit{A}'s integer power class, global parameters are better.

We also analyzed whether the ground state deformation and isospin correction factor affect the fitting of LDP parameters. Overall, the influence of ground state deformation on the coefficient fitting results is not significant. However, from the perspective of data fitting, whether to consider the influence of ground state deformation on the parameter fitting results is relatively large for nuclides in the transition region. 

In general, the experimental values of the cumulative number of atomic nuclear energy levels and the distance between neutron resonance energy levels near the neutron binding energy are obtained through the calculation of nuclear energy level density parameters. Therefore, in future work, we will use the programs of simplex method and composite method for fitting, and compare them with the LDP values obtained in this paper. It should be noted that when using the Olso method to extract LDP values, there may be a problem of fitting a larger LDP due to the large range of resonance energy of $\alpha$. This means that in the future, when using the Olso method to extract LDP values on a large scale, we still need to verify whether the fitted LDP satisfies the relationship of $a \propto A$.

\vspace{15pt}

The authors are grateful to Prof. Yuan Tian for fruitful discussions and suggestions. This work was supported by National Natural Science Foundation of China (Grants No. 12275115, No. 11447019).




\end{document}